\documentclass[
 reprint,
 prl,
superscriptaddress,
 amsmath,amssymb,
 aps,
 ]{revtex4-2}

\usepackage{graphicx}% Include figure files
\graphicspath{ {Figures/} }
\usepackage{dcolumn}% Align table columns on decimal point
\usepackage{bm}% bold math
\usepackage[hyperindex,breaklinks]{hyperref}% options so arXiv can wrap hrefs in references
\hypersetup{colorlinks=true}
\usepackage{breakurl}% so arXiv can wrap URLs in references
\usepackage[mathlines]{lineno}% Enable numbering of text and display math
\usepackage[caption=false]{subfig}
\usepackage{booktabs}
\usepackage{booktabs}
\usepackage{hhline}
\usepackage{graphicx}
\usepackage{float}  
\usepackage{xcolor} % for colored text
\usepackage[normalem]{ulem} % normalem keeps \emph from changing to underline
\usepackage{float}
\begin{document}
\setlength{\parindent}{1em}   % sets small indent
\setlength{\parskip}{0.0pt}     % removes vertical gap between paragraphs

\title{First Evidence of Solar Neutrino Interactions on $^{13}$C}

% SNO+ LaTeX Author List
\author{ M.\,Abreu }
\affiliation{ Laborat\'{o}rio de Instrumenta\c{c}\~{a}o e  F\'{\i}sica Experimental de Part\'{\i}culas (LIP), Av. Prof. Gama Pinto, 2, 1649-003, Lisboa, Portugal }
\affiliation{ Universidade de Lisboa, Instituto Superior T\'{e}cnico (IST), Departamento de F\'{\i}sica, Av. Rovisco Pais, 1049-001 Lisboa, Portugal }
\author{ A.\,Allega }
\affiliation{ Queen's University, Department of Physics, Engineering Physics \& Astronomy, Kingston, ON K7L 3N6, Canada }
\author{ M.\,R.\,Anderson }
\affiliation{ Queen's University, Department of Physics, Engineering Physics \& Astronomy, Kingston, ON K7L 3N6, Canada }
\author{ S.\,Andringa }
\affiliation{ Laborat\'{o}rio de Instrumenta\c{c}\~{a}o e  F\'{\i}sica Experimental de Part\'{\i}culas (LIP), Av. Prof. Gama Pinto, 2, 1649-003, Lisboa, Portugal }
\author{ D.\,M.\,Asner }
\affiliation{ SNOLAB, Creighton Mine \#9, 1039 Regional Road 24, Sudbury, ON P3Y 1N2, Canada }
\author{ D.\,J.\,Auty }
\affiliation{ University of Alberta, Department of Physics, 4-181 CCIS,  Edmonton, AB T6G 2E1, Canada }
\author{ A.\,Bacon }
\affiliation{ University of Pennsylvania, Department of Physics \& Astronomy, 209 South 33rd Street, Philadelphia, PA 19104-6396, USA }
\author{ T.\,Baltazar }
\affiliation{ Laborat\'{o}rio de Instrumenta\c{c}\~{a}o e  F\'{\i}sica Experimental de Part\'{\i}culas (LIP), Av. Prof. Gama Pinto, 2, 1649-003, Lisboa, Portugal }
\affiliation{ Universidade de Lisboa, Instituto Superior T\'{e}cnico (IST), Departamento de F\'{\i}sica, Av. Rovisco Pais, 1049-001 Lisboa, Portugal }
\author{ F.\,Bar\~{a}o }
\affiliation{ Laborat\'{o}rio de Instrumenta\c{c}\~{a}o e  F\'{\i}sica Experimental de Part\'{\i}culas (LIP), Av. Prof. Gama Pinto, 2, 1649-003, Lisboa, Portugal }
\affiliation{ Universidade de Lisboa, Instituto Superior T\'{e}cnico (IST), Departamento de F\'{\i}sica, Av. Rovisco Pais, 1049-001 Lisboa, Portugal }
\author{ N.\,Barros }
\affiliation{ Laborat\'{o}rio de Instrumenta\c{c}\~{a}o e  F\'{\i}sica Experimental de Part\'{\i}culas, Rua Larga, 3004-516 Coimbra, Portugal }
\affiliation{ Universidade de Coimbra, Departamento de F\'{\i}sica (FCTUC), 3004-516, Coimbra, Portugal }
% ...
\collaboration{ The SNO+ Collaboration }

\begin{abstract}
\newpage

The SNO+ Collaboration reports the first evidence of $^{8}\text{B}$ solar neutrinos interacting on $^{13}\text{C}$ nuclei. The charged current interaction proceeds through $^{13}\text{C} + \nu_e \rightarrow {}^{13}\text{N} + e^-$ which is followed, with a 10 minute half-life, by ${}^{13}\text{N} \rightarrow {}^{13}\text{C} + e^+ +\nu_e .$ The detection strategy is based on the delayed coincidence between the electron and the positron. Evidence for the charged current signal is presented with a significance of 4.2$\sigma$. Using the natural abundance of $^{13}\text{C}$ present in the scintillator, 5.7 tonnes of $^{13}\text{C}$ over 231 days of data were used in this analysis. The 5.6$^{+3.0}_{-2.3}$ observed events in the data set are consistent with the expectation of 4.7$^{+0.6}_{-1.3}$ events. This result is the second real-time measurement of CC interactions of $^{8}\text{B}$ neutrinos with nuclei and constitutes the lowest energy observation of neutrino interactions on $^{13}\text{C}$ generally. This enables the first direct measurement of the CC $\nu_e$ reaction to the ground state of ${}^{13}\text{N}$, yielding an average cross section of $(16.1 ^{+8.5}_{-6.7} (\text{stat.}) ^{+1.6}_{-2.7} (\text{syst.}) )\times 10^{-43}$~cm$^{2}$ over the relevant $^{8}\text{B}$ solar neutrino energies.

\end{abstract}

\maketitle
\textit{Introduction --- }
The SNO+ liquid scintillator experiment reports the first evidence of the charged current (CC) interaction of solar neutrinos with $^{13}$C, ${}^{13}\text{C} + \nu_e \rightarrow {}^{13}\text{N} + e^-$. The interaction has a threshold energy of 2.2 MeV, and is detected using the delayed coincidence between the prompt electron and the $^{13}$N, which undergoes $\beta^{+}$ decay with a half-life of 9.96 minutes and a Q-value of 2.2 MeV.

The flux and spectrum of $^{8}$B electron neutrinos from the Sun have been constrained by measurements of CC and neutral current (NC) neutrino interactions on various targets, including elastic scattering (ES) interactions on electrons, from numerous independent solar neutrino experiments \cite{Solar_Annual_Review}. Unlike NC or ES reactions, CC interactions directly relate the outgoing-electron energy to the incident-neutrino energy, providing a sensitive probe of the $^{8}$B spectrum.  Results are wholly consistent with the Standard Solar Model together with neutrino oscillations \cite{Solar_models}, with oscillation parameters also confirmed by terrestrial experiments \cite{Oss_Annual_Review}. Given the solar electron-neutrino spectrum established by this framework, we are now able to measure the CC cross section on a different nucleus, $^{13}\text{C}$, for the first time.

\vspace{+0.4em}
\textit{The SNO+ Detector --- }
The SNO+ detector is located 2 km underground at the SNOLAB facility in the Creighton Mine near Sudbury, Canada. SNO+ inherits much of the infrastructure from the SNO experiment \cite{SNO_Experiment}, including a spherical acrylic vessel (AV) of 6 m radius located within a barrel-shaped cavity excavated in the rock. During the data-taking for this analysis, the vessel was filled with 780 tonnes of linear alkylbenzene (LAB) doped with 2.2 g/L of 2,5-diphenyloxazole (PPO) \cite{Anderson_2021}. A secondary fluor, 1,4-bis(2-methylstyryl)benzene (bis-MSB), was subsequently added to the scintillator to boost light collection. The scintillator cocktail is viewed by $\sim$9300 photomultiplier tubes (PMTs) supported by a geodesic stainless steel structure (PSUP) of approximately 8.9~m radius. The volume between the AV and the cavity walls, including the space between the AV and PSUP, is filled with approximately 7000 tonnes of ultra-pure water, which shields the scintillator from the radioactivity in the rock (cavity walls) and the PMT array. In addition, 91 PMTs are mounted outward-looking (OWL) from the PSUP to detect light originating from the surrounding water, such as that produced by muons. A detailed description of the SNO+ detector is given in \cite{Albanese_2021}. A flat overburden of 2070~m (6010 meters water equivalent) provides a shield against cosmic muons, giving a muon rate of (0.286 ± 0.009) $\mu$/m${}^2$/d \cite{SNO_Muon_Flux}. The depth of SNOLAB was critical to the current measurement, suppressing muon-induced backgrounds to a negligible level.

\vspace{+0.4em}
\textit{Data --- } 
This analysis uses data collected between 04\textsuperscript{th} May 2022 and 29\textsuperscript{th} June 2023, and has a livetime of 247.3 days. During this period, the measured light yield was $\sim$210~PMT hits/MeV. The trigger threshold was $\sim$23~PMT hits (within an event window of 400 ns), corresponding to $\sim$0.11~MeV, well below the energy of the $^{13}$C signal of interest. The analysed data runs are selected based on data quality and detector performance criteria, and event-level selection criteria are applied to remove instrumental backgrounds~\cite{SNO+_nucleon_decay}.

The properties of particle interactions are inferred using the times and locations of hit PMTs. The photon time-of-flight is used to reconstruct the interaction position. The number of hit PMTs is approximately proportional to the energy a particle deposits. The resolution of the reconstructed energy $E$ is $\sim$6.5\%/$\sqrt{E[MeV]}$ for an electron at the centre of the detector for this dataset. The reconstructed position resolution for a 2.5~MeV electron at the centre is 12~cm in each Cartesian axis. 

\vspace{+0.4em}
\textit{Signal --- }
There are two possible interactions of neutrinos with $^{13}$C nuclei \cite{ARAFUNE1989186}: a CC interaction with a threshold of 2.2 MeV and a NC interaction starting at 3.1~MeV. As discussed in \cite{IANNI200538}, the delayed coincidence signal associated with the CC interaction provides a powerful tool for identifying these interactions. The CC interaction proceeds through ${}^{13}\text{C} + \nu_e \rightarrow {}^{13}\text{N} + e^-$   with a Q-value of 2.2~MeV.  $^{13}$N then $\beta^{+}$ decays with a half-life of $t_{1/2} = 9.96$~minutes, to $^{13}$C: ${}^{13}\text{N} \rightarrow {}^{13}\text{C} + e^+ +\nu_e$. Only interactions to the $^{13}$N nuclear ground state were considered, as $^{13}$N excited states decay via proton emission too rapidly to produce a viable delayed coincidence signal {\cite{N13_excited_states}. The expected rate of $\nu_e$ -- $^{13}$C interactions inside SNO+ is calculated using:
\begingroup
\setlength{\abovedisplayskip}{10pt}   % space above
\setlength{\belowdisplayskip}{5pt}   % space below
\[ \text{R}_{^{13}\text{C}}= \text{N}_{^{13}\text{C}}\Phi_{^{8}\text{B}}\int_{}^{}\sigma(E_{\nu})S_{\nu}(E_{\nu})P_{ee}(E_{\nu})dE_{\nu}\]
\endgroup
\noindent
where $\text{R}_{^{13}\text{C}}$ is the rate of interaction; $\text{N}_{^{13}\text{C}} = (3.8^{+0.3}_{-0.6})\times 10^{29}$ the number of $^{13}$C atoms (target nuclei) inside the SNO+ detector;  $\Phi_{^{8}\text{B}} = (5.2 \pm  0.1)\times 10^{6}~\text{cm}^{-2}\text{s}^{-1}$ is the $^{8}$B solar neutrino flux \cite{B8_Flux}; $\sigma(E_{\nu})$ is the interaction cross section as a function of the incoming neutrino energy~($E_{\nu}$) \cite{Cross_Section_2012}; $S_{\nu}(E_{\nu})$ is the normalised $^{8}$B energy spectrum \cite{B8_spectrum}; and $P_{ee}(E_{\nu})$ is the averaged MSW electron neutrino survival probability assuming the adiabatic approximation using the globally fit neutrino oscillation parameters \cite{Neutrinos_oscillation_parameters_2024}. The expected interaction rate is $21.9^{+2.0}_{-2.7}$~ev/yr/kT of LAB. Uncertainty in the expected rate is driven by the variation in the natural abundance of ${}^{13}\text{C: } 1.1^{+0.06}_{-0.14}$\% \cite{C13_Natural_Abundance}. 

\vspace{+0.4em}
\textit{Analysis Methods --- }
An energy threshold of 5~MeV on the prompt event removes almost all background from internal radioactive decays, such as the $\beta^-$ decay of $^{208}$Tl (Q-value = 5.0~MeV). A 15 MeV prompt upper cut was applied, covering the endpoint of the $^{8}$B solar neutrino energy spectrum. With this prompt event selection, the background to the ${}^{13}\text{C}$ signal is expected to consist almost exclusively of accidental coincidences between ${}^{8}\text{B}$ solar neutrino electron scattering events and unrelated radioactive decays that match the delayed signal criteria. A blinded approach was used, where the delayed data events were not identified until after the cuts and analysis methods were finalised.

The 10 minute half-life of the $^{13}$N presents a challenge for accidental coincidences, as many delayed events may be associated with a given prompt event due to the relatively high rate ($\sim$0.02 Hz) of background events in the delayed energy window. Additional data selection cuts were therefore introduced and tuned to optimise the expected signal-to-noise ratio. A Fiducial Volume (“FV”) cut was used to eliminate dominant external backgrounds near the detector's edge. The “In-Time-Ratio” (ITR) describes the ratio of the number of hit PMTs within a 7.5 ns “prompt time window” to all hit PMTs in an event, and is used to eliminate residual instrumental backgrounds. The allowed ITR region was separately optimised for the expected prompt and delayed signals, relative to their respective backgrounds. 

A lower energy cut of 1.0~MeV for delayed events was chosen, as the rate of radioactive backgrounds increases below this, such as the $\beta^-$ decay of \textsuperscript{210}Bi (Q-value = 1.16~MeV). The upper energy cut for the delayed event was defined as 2.2~MeV to match the Q-value for the delayed $^{13}\text{N}$ decay. Wide selection windows for the difference in time $(\Delta T)$ and vertex position reconstructions $(\Delta R)$ between the prompt and delayed events, capture more than 98\% of the ${}^{13}\text{C}$ signal. The full set of optimised data selection cuts for the prompt and delayed events are shown in Table \ref{tab:selection_cuts}.

\begin{table}[h]
\centering
\setlength{\tabcolsep}{15pt}
\begin{tabular}{@{}ccc@{}}
\toprule
& \textbf{Prompt}      & \textbf{Delayed}             \\ 
\midrule
$R$ (m)           & $<$ 5.3      & $<$ 5.3                 \\
ITR               & 0.1 -- 0.35    & 0.125 -- 0.3        \\
Energy (MeV)      & 5.0 -- 15.0    & 1.0 -- 2.2          \\ 
$\Delta R$ (m)    & \multicolumn{2}{c}{ $<$ 1.0}        \\
$\Delta T$ (minutes)   & \multicolumn{2}{c}{0.01 -- 60.0} \\
\bottomrule
\end{tabular}
\caption{Initial selection cuts for prompt and delayed events.}
\label{tab:selection_cuts}
 \vspace{-2.0em}
\end{table}
\noindent

The expected accidental coincidence rate was obtained by multiplying the prompt candidate rate by the probability of an accidental coincidence. The latter was evaluated with a data-driven method: random positions and times were sampled from the SNO+ dataset, the fraction of these random events coincident with real events passing the delayed cuts was measured.

The efficiency for selecting signal events was determined using Monte Carlo (MC) simulations using the GEANT-4 based SNO+ RAT simulation package \cite{Albanese_2021}. The final signal efficiency was 44$^{+2.6}_{-1.6}\%$ and the expected rate of random backgrounds was 0.34 per prompt event. The total number of prompt events was 173.  The uncertainty on the MC efficiency is dominated by the estimated FV uncertainty. The MC systematics were largely constrained using coincident $^{214}$Bi -- $^{214}$Po decays from detector intrinsic background $^{222}$Rn levels. The high Q-value of $^{214}$Bi $\beta^{-}$ decay (3.27~MeV), followed by the short half-life of $^{214}$Po (164~$\mu$s), enables the selection of a highly pure sample of $^{214}$Bi -- $^{214}$Po events in data. This sample was compared with MC to evaluate systematic differences, and the resulting corrections and uncertainties were then propagated to the $^{13}$C signal MC.}

\vspace{+0.4em}
\textit{Correlated Backgrounds --- }
Any correlated signal in the detector that matches the characteristics of the $^{13}\text{C}$ signal would present a background to this measurement. Delayed coincidences occurring in the decay chains of natural radioactivity were removed with the prompt energy cut of $>$ 5 MeV. Inverse beta decay (IBD) events from reactor antineutrino interactions could cause a correlated signal as the prompt reactor and delayed neutron capture energy spectrum partly overlap with the ranges used in this analysis. However, due to the short capture time of the neutron ($\sim$200$~\mu$s), applying the lower $\Delta T$ cut of 0.01 minutes reduces this background to a negligible level. There are two remaining expected correlated backgrounds: cosmogenic spallation backgrounds and atmospheric neutrinos.

Muon spallation can produce a multitude of short-lived radioisotopes, thus creating possible delayed coincidences between the muon and the subsequent decay of the spallation daughter, or between two spallation product decays. Muons in the data set are identified using two methods: if the total number of PMTs hit exceeds 3750, or if 5 or more OWL PMTs are hit while the total PMT hits is $\geq$ 750. Following each identified muon, to suppress spallation backgrounds, events were vetoed for 60~seconds in the prompt selection and 20~seconds in the delayed selection. The dominant cosmogenic background expected after the vetoes are from $^{11}$Be (Q-value = 11.5~MeV, $t_{1/2}$ = 13.8~seconds) and $^{11}$C (Q-value~=~2.0~MeV, $t_{1/2}$ = 20.3~minutes).  The high Q-value of $^{11}$Be means that it cannot be removed with an energy cut without significant signal sacrifice. $^{11}$C decays via positron emission, similarly to $^{13}$N, and has a relatively long half-life. Thus, a muon producing both isotopes, a prompt $^{11}$Be decay and a delayed $^{11}$C decay, constitutes a correlated background for the $^{13}$C signal. The expected number of $^{11}$Be and $^{11}$C events can be calculated by scaling the rates from KamLAND \cite{KamLAND_muon_spallation}; after the time vetoes this gives a negligible expected rate of 0.0015 ev/yr/kT of LAB in SNO+. The livetime loss due to both muon vetoes is 6.80\%, giving a total livetime of 230.51~days.

Atmospheric neutrino interactions generally do not involve a coincidence, or occur on the microsecond time scale, and are therefore negligible within the time window for this analysis. The NC interaction on $^{12}$C can produce isotopes that mimic the delayed signal, such as $^{11}$C: $ (\nu_x + {}^{12}\text{C} \rightarrow  \nu_x + n + {}^{11}\text{C})$. 
The thermalisation of the high-energy neutron, together with the subsequent decay of $^{11}$C, can generate a correlated background to the $^{13}$C signal. By looking for the characteristic 2.2 MeV gamma ray from neutron capture within 4 ms of a candidate prompt event, this background can be identified, reducing it to a level such that the total expected atmospheric rate is $< 0.01$~ev/yr/kT~of LAB.

\begin{figure}[H]
\centering
\includegraphics[width=1.0\linewidth]{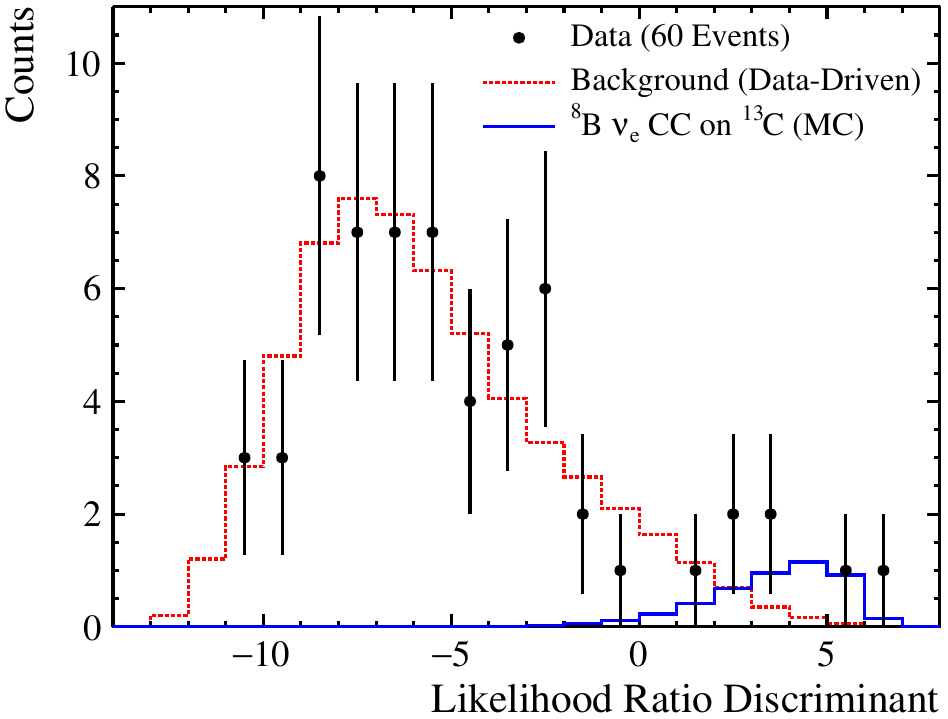}
\caption{The LR distribution for data (black) compared to the expected likelihood LR for the signal (blue) and background (red dashed).}
\label{fig:fig1}
\vspace{-1.0em}
\end{figure}

\vspace{+0.4em}
\textit{Results --- }
Signal and background events were differentiated using a likelihood ratio (LR), based on the $\Delta T$, $\Delta R$ and delayed event energy \cite{SuppMat}. Inclusion of the delayed-energy PDF enables separation between the high-background region ($<$~1.2 MeV) and the low-background region ($>$~1.2 MeV). The LR distribution from data is shown in Figure \ref{fig:fig1}, with the signal PDF scaled using the theoretical rate and the signal efficiency and the accidental background scaled to the expected rate determined through the data-driven procedure described earlier. 

\begin{figure}[htbp]
\centering
\includegraphics[width=1.0\linewidth]{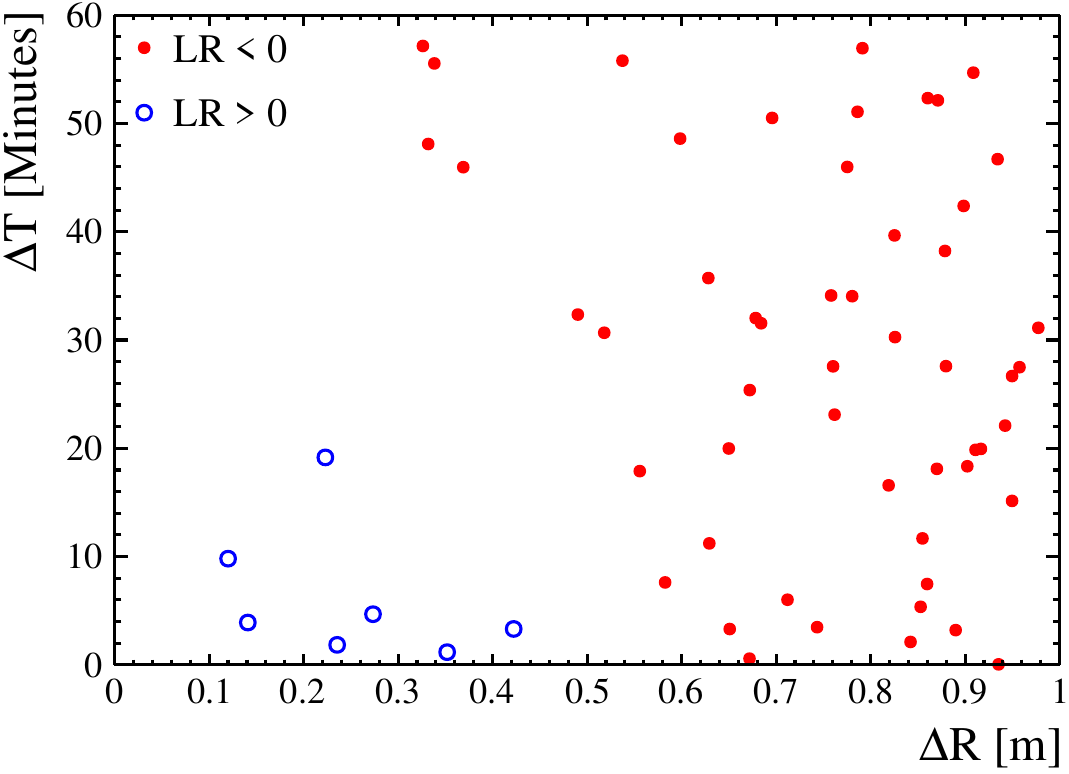}
\caption{Distribution of the difference in time $(\Delta T)$ and 3D position $(\Delta R)$ between the prompt and delayed events. Signal-like (LR $> 0$) events are shown as open blue circles and background-like (LR $< 0$) events as filled red circles.}
\label{fig:fig2}
\end{figure}

There were 63.2 events expected, which agrees with the observed 60 events. Furthermore, the background distribution closely matches the data, including the fraction of prompt events with more than one delayed event candidate. Events with a LR $> 0$ show more consistency with the $^{13}$C solar neutrino signal. Figure \ref{fig:fig2} shows the $\Delta R$ and $\Delta T$ distribution of all events. Events with low $\Delta R$ and $\Delta T$,  shown in open blue circles, correspond to the signal candidate events with LR $> 0$. The background events, shown with filled red circles, have a uniform time distribution, as expected from a random background. 

\begin{figure}[htbp]
\centering
\includegraphics[width=1.0\linewidth]{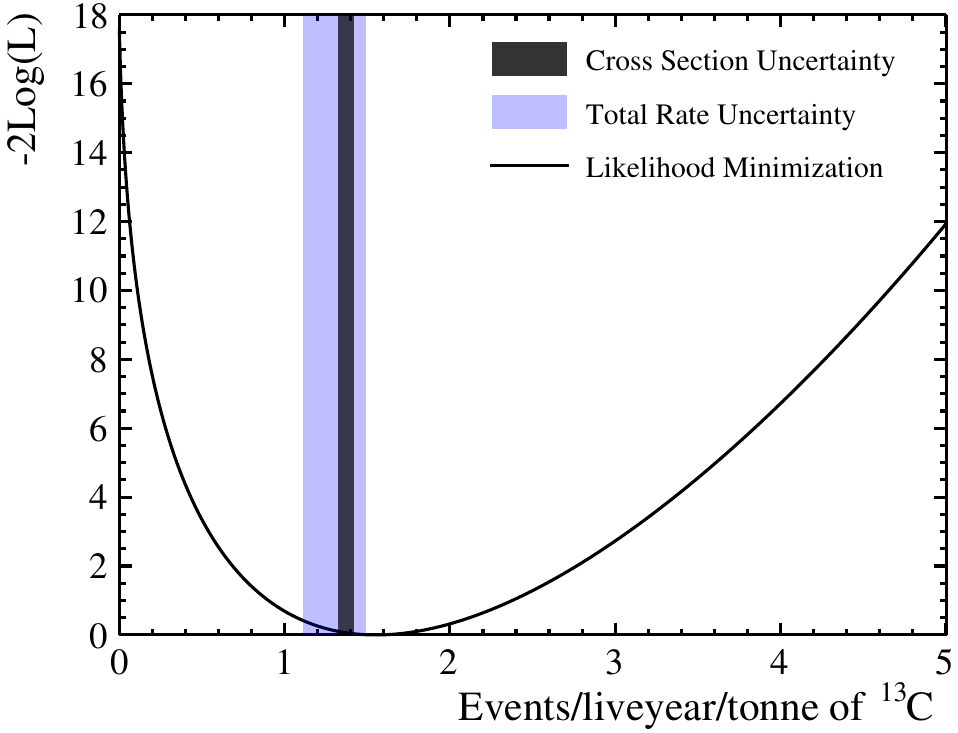}
\caption{\label{fig:figure} The observed likelihood distribution for the $^{13}\text{C}$ interaction rate, compared to the expected rate. The theoretical prediction (shaded blue) agrees with the best fit signal rate of 1.6$^{+0.82}_{-0.65}$ ev/yr/tonne of $^{13}$C. The prediction rate only considering cross section uncertainty is shown in shaded black.}
\label{fig:fig3}
\end{figure}

The likelihood PDFs can be used to construct a mixed signal and background model, from which the overall relative likelihood can be plotted as a function of signal content, as shown in Figure \ref{fig:fig3}. The background-only null hypothesis is rejected at 4.2$\sigma$, and the best fit number of signal events is 1.6$^{+0.82}_{-0.65}$ ev/yr/tonne of $^{13}$C, consistent with the expectation of 1.3$^{+0.17}_{-0.21}$ ev/yr/tonne of $^{13}$C. The uncertainties on the measured signal rate are derived from the likelihood minimisation using Wilks' theorem. The uncertainty in the expected rate is dominated by the 9\% average uncertainty on the natural isotopic abundance of $^{13}$C. 

\begin{figure}[htbp]
\centering
\includegraphics[width=1.0\linewidth]{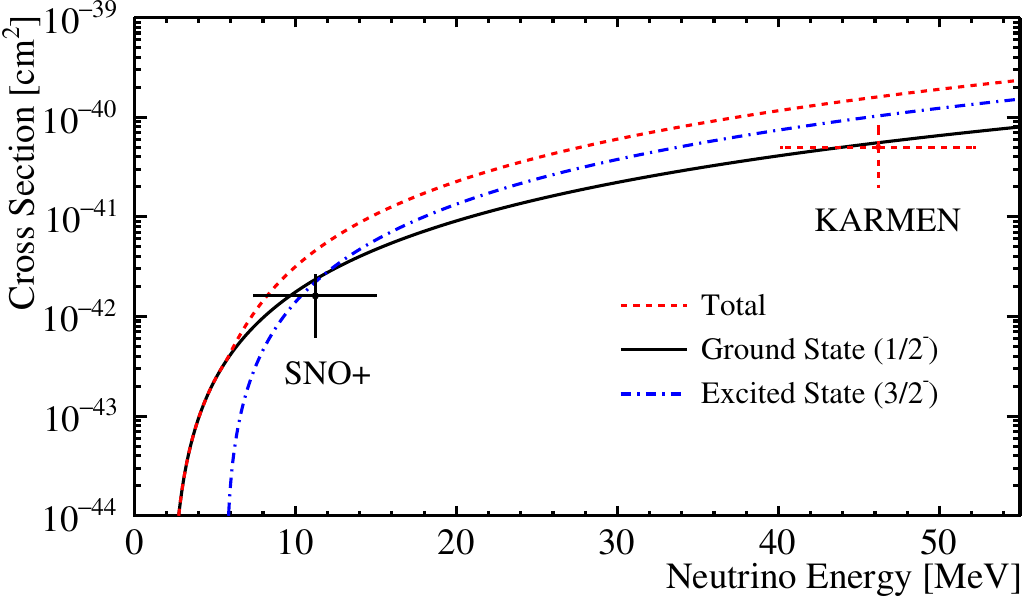}
\caption{\label{fig:figure}
The SNO+ measured cross section for the $\nu$ + $^{13}$C CC reaction (black point) compared to the theoretical prediction \cite{Cross_Section_2012} for the ground state interaction (black line). Also shown is the previous KARMEN result \cite{KARMEN} (red point) compared to the theoretical total cross section (red dashed). The theoretical cross section for the excited state interaction (blue dot-dash) is also shown. The horizontal error bars show the range of neutrino energies used for each measurement, with the data points placed in the centre of that range.
}
\label{fig:fig4}
\end{figure}

Finally, the observed rate of interactions can be combined with previous measurements of the ${}^{8}\text{B}$ flux, the global best fit neutrino oscillation parameters and the number of ${}^{13}\text{C}$ atoms to determine the interaction cross section. The flux-weighted average cross section is $(16.1 ^{+8.5}_{-6.7} (\text{stat.}) ^{+1.6}_{-2.7} (\text{syst.}) )\times 10^{-43}$~cm$^{2}$ for $^{8}$B neutrino events above 7.2 MeV. The only other $^{13}$C neutrino cross~section measurement comes from the KARMEN experiment \cite{KARMEN}. As KARMEN was operating at higher energies and was not looking for the delayed coincidence, they were sensitive to the sum of all interaction channels, not just the ground state. The results presented here, therefore, constitute the first specific measurement of the ground state cross section for this reaction. The cross sections are shown in Figure \ref{fig:fig4} together with the theoretical predictions \cite{Cross_Section_2012}. The 13\% systematic uncertainty on the measured cross section is subdominant to the 47\% statistical uncertainty.

\newpage
\vspace{+0.4em}
\textit{Summary ---}
This paper presents the first evidence of $^{8}$B solar neutrinos interacting on $^{13}$C at a significance level of 4.2$\sigma$ using 231 days of data. Using the correlation of the spatial and temporal signals between the prompt electron and the delayed decay of the $^{13}$N, the signal can be separated from accidental backgrounds. The observed number of 5.6$^{+3.0}_{-2.3}$ events, is consistent with the expected 4.7$^{+0.6}_{-1.3}$ events. The derived cross section for $^{8}$B neutrino events above 7.2 MeV is $(16.1 ^{+8.5}_{-6.7} (\text{stat.}) ^{+1.6}_{-2.7} (\text{syst.}) )\times 10^{-43}$~cm$^{2}$, in agreement with theoretical predictions. This constitutes both the lowest energy observation of neutrino interactions on $^{13}\text{C}$, and the first direct cross section measurement for the interaction to the $^{13}$N ground state. The measured low-energy cross section further serves as a benchmark for theoretical studies employing state-of-the-art models of nuclear dynamics and advanced computational techniques, such as the ab initio quantum MC approaches reviewed in \cite{Quantum_MC_Carlson}.

\vspace{+1.0em}
\begin{acknowledgments}
\textit{Acknowledgments --- }Capital funds for SNO\raisebox{0.5ex}{\tiny\textbf{+}} were provided by the Canada Foundation for Innovation and matching partners: 
Ontario Ministry of Research, Innovation and Science, 
Alberta Science and Research Investments Program, 
Queen’s University at Kingston, and 
the Federal Economic Development Agency for Northern Ontario. 
This research was supported by 
{\it Canada: }
the Natural Sciences and Engineering Research Council of Canada, 
the Canadian Institute for Advanced Research, 
the Ontario Early Researcher Awards, 
the Arthur B. McDonald Canadian Astroparticle Physics Research Institute; 
{\it U.S.: }
the Department of Energy (DOE) Office of Nuclear Physics, 
the National Science Foundation and the DOE National Nuclear Security
Administration through the Nuclear Science and Security Consortium; 
{\it UK: }
the Science and Technology Facilities Council and the Royal Society; 
{\it Portugal: } 
Funda\c{c}\~{a}o para a Ci\^{e}ncia e a Tecnologia (FCT-Portugal); 
{\it Germany: }
the Deutsche Forschungsgemeinschaft; 
{\it Mexico: }
DGAPA-UNAM and Consejo Nacional de Ciencia y Tecnolog\'{i}a; 
{\it China: }
the Discipline Construction Fund of Shandong University.  
We also thank SNOLAB and SNO\raisebox{0.5ex}{\tiny\textbf{+}} technical
staff; the Digital Research Alliance of Canada; the
GridPP Collaboration and support from Rutherford
Appleton Laboratory; and the Savio computational cluster
at the University of California, Berkeley. Additional long-term
storage was provided by the Fermilab Scientific Computing
Division.

\vspace{0.4em}
\textit{Data availability --- }For the purposes of open access, the authors have applied a Creative Commons Attribution licence to any Author Accepted Manuscript version arising. Representations of the data relevant to the conclusions drawn here are provided within this paper and its supplemental material \cite{SuppMat}.
\end{acknowledgments}

\newpage
% \bibliographystyle{apsrev}
% \bibliography{References}
\providecommand{\noopsort}[1]{}\providecommand{\singleletter}[1]{#1}%

\end{document}

%%%%%%%%%%%%%%%%%%%%%%%%%%%%%%%%%%%%%%%%%%%%%%%%%%%%%%%%%%%%%%%%%%%%%%%%%%%%%%%%%%%%%%%%%
%%%%%   %%%%%
%%%%%   Supplemental Material for the   %%%%%
%%%%%   SNO+ Analysis of Solar Neutrinos on Carbon 13    %%%%%
%%%%%   Prepared by Gulliver Milton (2025)    %%%%%
%%%%%   %%%%%
%%%%%%%%%%%%%%%%%%%%%%%%%%%%%%%%%%%%%%%%%%%%%%%%%%%%%%%%%%%%%%%%%%%%%%%%%%%%%%%%%%%%%%%%%

{\centering{}
\LARGE{SNO+ Experiment} \\ 
\Large{Supplemental Material} \\
}

\bigskip
\bigskip
\normalsize
This document provides supplemental information relevant for the paper ``First Evidence of Solar Neutrino Interactions on $^{13}$C''. Where reference is made to x,y,z coordinates, these are defined with the origin at the centre of the AV and the z axis passing through the neck of the detector.

\begin{table}[htbp]
\setlength{\tabcolsep}{5pt} 
\renewcommand{\arraystretch}{1.05} 
\centering
\begin{tabular}{@{}cccc|cccc|c@{\hspace{10pt}}c@{\hspace{10pt}}c@{\hspace{10pt}}c@{}}
\toprule
\toprule
\multicolumn{4}{c|}{\textbf{Prompt}} & \multicolumn{4}{c|}{\textbf{Delayed}} & \multirow{2}{*}{$\Delta T$ [min]} & \multirow{2}{*}{$\Delta R$ [m]} & \multirow{2}{*}{LR} & \multirow{2}{*}{Date} \\
E [MeV] & x [m] & y [m] & z [m] & E [MeV] & x [m] & y [m] & z [m] & & & & \\
\midrule

7.71   & -3.18   & -0.22   & 3.92   & 1.02   & -2.49   & 0.03   & 4.35   & 2.23   & 0.850   & -6.43    & 05/05/2022    \\
---   & ---   & ---    & ---    & 1.03   & -2.67   & 0.07   & 3.70   & 35.75   & 0.625   & -5.71    & 05/05/2022    \\
5.59   & 1.63   & -0.37   & 1.92   & 1.12   & 1.32   & 0.48   & 2.01   & 18.24   & 0.909   & -7.30    & 09/05/2022    \\
9.55   & 0.15   & -0.98   & 5.16   & 1.04   & 0.62   & -1.09   & 4.95   & 30.99   & 0.519   & -3.49    & 17/05/2022    \\
---   & ---    & ---    & ---    & 1.03   & 0.01   & -1.61   & 4.98   & 31.91   & 0.671   & -6.06    & 17/05/2022    \\
5.12   & -0.99   & -0.82   & -3.70   & 1.08   & -0.75   & -0.69   & -3.52   & 48.13   & 0.331   & -1.21    & 21/05/2022    \\
6.03   & 1.78   & 1.51   & -1.64   & 1.08   & 2.00   & 0.84   & -1.07   & 54.62   & 0.901   & -10.30   & 27/05/2022    \\
8.77   & 2.25   & 2.77   & -0.09   & 1.03   & 1.64   & 2.68   & -0.69   & 7.45   & 0.858   & -7.23    & 30/05/2022    \\
---    & ---    & ---    & ---    & 1.07   & 1.84   & 2.44   & -0.65   & 27.18   & 0.767   & -6.86    & 30/05/2022    \\
5.42   & -2.14   & 1.34   & 0.39   & 1.11   & -2.54   & 1.34   & -0.38   & 52.04   & 0.872   & -8.75    & 13/06/2022    \\
8.14   & 3.66   & -3.25   & 1.45   & 1.07   & 3.18   & -3.38   & 1.39   & 31.91   & 0.496   & -2.56    & 20/06/2022    \\
---    & ---    & ---   & ---    & 1.47   & 3.56   & -3.31   & 1.54   & 3.96   & 0.145   & 6.02   & 20/06/2022    \\
5.89   & 0.14   & -0.83   & 5.19   & 1.07   & 0.03   & -1.64   & 4.89   & 27.29   & 0.872   & -7.70    & 20/06/2022    \\
5.20   & -1.64   & -2.34   & -4.01   & 1.08   & -2.10   & -1.92   & -3.84   & 19.85   & 0.644   & -4.31    & 22/06/2022    \\
7.09   & -3.84   & -2.72   & 1.64   & 1.04   & -3.64   & -3.00   & 0.88   & 30.22   & 0.830   & -8.08    & 17/07/2022    \\
9.91   & -2.08   & -3.68   & -2.99   & 1.07   & -2.08   & -3.93   & -2.25   & 46.04   & 0.779   & -8.03    & 10/08/2022    \\
5.14   & 0.23   & -0.46   & 3.71   & 1.11   & 0.53   & 0.05   & 4.03   & 31.35   & 0.682   & -5.29    & 22/08/2022    \\
6.42   & 1.06   & -2.31   & -1.09   & 1.00   & 1.38   & -2.60   & -1.74   & 33.81   & 0.781   & -8.35    & 29/08/2022    \\
6.22   & -1.54   & 1.81   & 2.63   & 1.44   & -1.34   & 2.47   & 2.81   & 6.50   & 0.711   & -2.34    & 06/09/2022    \\
5.76   & 0.33   & -0.93   & 5.06   & 1.15   & -0.04   & -1.62   & 4.61   & 3.31   & 0.893   & -5.36    & 11/09/2022    \\
---    & ---    & ---    & ---    & 1.11   & -0.16   & -1.34   & 4.89   & 3.58   & 0.652   & -2.85    & 11/09/2022    \\
5.67   & 0.45   & 1.12   & -2.81   & 1.07   & 0.52   & 1.33   & -2.53   & 45.90   & 0.361   & -1.47    & 12/09/2022    \\
6.53   & -3.41   & 1.76   & -0.08   & 1.36   & -2.97   & 2.37   & 0.05   & 23.13   & 0.766   & -3.05    & 25/09/2022    \\
9.30   & -3.19   & -1.95   & -1.40   & 1.09   & -3.51   & -2.39   & -1.49   & 17.95   & 0.55    & -2.06    & 30/09/2022    \\
5.92   & -2.32   & -3.79   & 0.08   & 1.14   & -2.16   & -3.97   & 0.21   & 4.60   & 0.275   & 3.00   & 03/10/2022    \\
---    & ---    & ---    & ---    & 1.01   & -2.24   & -3.72   & -0.23   & 57.52   & 0.323   & -2.61    & 03/10/2022    \\
5.85   & 2.72   & 1.21   & 3.01   & 1.73   & 2.66   & 1.04   & 2.21   & 16.64   & 0.816   & -4.03    & 08/10/2022    \\
5.76   & -3.05   & 0.55   & 3.74   & 1.01   & -3.31   & 1.00   & 3.48   & 7.31   & 0.587   & -3.51    & 29/10/2022    \\
9.17   & -5.06   & -0.77   & 1.02   & 1.63   & -4.61   & -1.59   & 1.17   & 26.62   & 0.947   & -6.48    & 30/10/2022    \\
10.15   & -0.22   & 1.28   & -1.98   & 1.34   & -0.32   & 1.83   & -2.74   & 15.34   & 0.944   & -5.38    & 05/11/2022    \\
7.31   & -1.13   & -2.23   & -3.99   & 1.13   & -1.22   & -2.23   & -3.90   & 9.72   & 0.129   & 3.49   & 17/11/2022    \\
7.91   & 2.06   & -2.63   & -3.47   & 1.79   & 2.52   & -1.99   & -4.05   & 30.93   & 0.976   & -7.41    & 21/11/2022    \\
8.50   & 2.92   & 3.53   & -0.26   & 1.60   & 3.20   & 4.14   & 0.27   & 11.63   & 0.857   & -3.78    & 26/11/2022    \\
12.95   & 0.27   & -1.12   & 4.09   & 1.00   & 0.10   & -1.00   & 4.66   & 48.95   & 0.599   & -6.36    & 04/12/2022    \\
9.43   & 0.62   & 2.85   & -0.40   & 1.35   & 1.31   & 2.43   & -0.93   & 27.33   & 0.959   & -6.98    & 19/12/2022    \\
8.89   & 0.67   & 0.00   & -0.20   & 1.11   & 0.89   & -0.26   & 0.39   & 0.52   & 0.672   & -3.24    & 23/12/2022    \\
7.52   & 3.80   & 0.97   & 0.98   & 1.00   & 3.76   & 1.29   & 0.90   & 55.71   & 0.339   & -2.80    & 10/01/2023    \\
---    & ---    & ---    & ---    & 1.01   & 3.21   & 0.87   & 0.37   & 5.21   & 0.852   & -6.81    & 10/01/2023    \\
7.31   & 0.50   & 0.37   & 4.71   & 1.01   & 0.16   & 1.16   & 4.33   & 46.56   & 0.937   & -10.92   & 18/01/2023    \\
---    & ---    & ---    & ---    & 1.04   & 0.50   & 0.11   & 3.97   & 51.27   & 0.788   & -9.86    & 18/01/2023    \\
8.35   & 0.68   & -1.65   & -2.44   & 1.04   & 1.23   & -1.63   & -1.82   & 39.74   & 0.825   & -8.88    & 20/01/2023    \\
6.02   & -3.54   & -2.85   & 2.18   & 1.03   & -3.17   & -3.27   & 1.52   & 52.33   & 0.868   & -10.23   & 20/01/2023    \\
5.90   & 0.22   & -0.85   & 4.29   & 1.11   & 0.87   & -0.77   & 4.65   & 3.48   & 0.746   & -4.05    & 22/01/2023    \\
5.95   & 3.56   & -2.52   & -0.96   & 1.02   & 3.72   & -3.12   & -1.58   & 38.09   & 0.877   & -9.33    & 23/01/2023    \\
9.90   & 0.60   & -1.03   & -4.98   & 1.08   & 1.01   & -0.51   & -4.86   & 25.78   & 0.675   & -5.39    & 23/01/2023    \\
---    & ---    & ---    & ---    & 1.01   & 0.17   & -0.82   & -5.22   & 56.08   & 0.539   & -5.79    & 23/01/2023    \\
5.77   & -2.50   & -1.68   & -1.20   & 1.01   & -3.17   & -1.50   & -1.51   & 33.89   & 0.758   & -8.18    & 31/01/2023    \\
6.97   & 0.28   & 0.85   & 5.00   & 1.06   & -0.32   & 1.50   & 5.06   & 42.17   & 0.891   & -9.23    & 10/02/2023    \\
8.44   & -2.09   & -4.10   & -1.51   & 1.02   & -1.33   & -4.30   & -0.99   & 22.03   & 0.943   & -8.93    & 17/02/2023    \\
5.14   & -0.16   & 1.90   & 1.68   & 2.07   & 0.25   & 1.59   & 2.43   & 19.97   & 0.916   & -8.66    & 24/02/2023    \\
---    & ---    & ---    & ---    & 1.04   & -0.89   & 2.11   & 1.23   & 18.01   & 0.875   & -7.58    & 24/02/2023    \\
6.57   & -0.67   & 1.16   & -1.61   & 1.13   & -0.97   & 0.61   & -2.11   & 56.66   & 0.80    & -7.97    & 01/03/2023    \\ 
9.26   & -0.77   & -1.75   & -1.21   & 1.05   & -1.51   & -1.36   & -1.62   & 0.26   & 0.932   & -7.30    & 07/03/2023    \\
---    & ---    & ---   & ---    & 1.18   & -0.58   & -1.86   & -1.21   & 19.16   & 0.223   & 2.99   & 07/03/2023    \\
5.16   & 1.90   & 0.16   & -3.08   & 1.48   & 2.27   & 0.90   & -3.46   & 20.14   & 0.913   & -5.06    & 08/03/2023    \\
8.09   & 0.18   & 3.49   & -1.26   & 1.04   & 0.39   & 3.46   & -1.54   & 0.68   & 0.353   & 1.41   & 08/03/2023    \\
5.48   & 2.89   & 0.19   & 1.64   & 1.71   & 3.31   & 0.01   & 1.21   & 11.35   & 0.625   & -0.03    & 28/05/2023    \\
6.95   & -0.18   & -4.16   & -2.61   & 1.85   & -0.72   & -3.90   & -2.25   & 50.97   & 0.698   & -4.84    & 21/06/2023    \\
---    & ---    & ---    & ---    & 1.66   & -0.15   & -3.98   & -2.47   & 1.84   & 0.232   & 5.32   & 21/06/2023    \\
9.84   & -0.47   & 3.62   & 3.03   & 1.45   & -0.40   & 3.30   & 3.31   & 3.15   & 0.427   & 3.59   & 21/06/2023    \\
\bottomrule
\bottomrule
\end{tabular}
\caption{Information about the 60 coincidence pairs selected. The dashed lines indicate that the same prompt event had multiple delayed events.}
\label{tab:prompt_delayed}
\end{table}

\clearpage

\section*{Event distributions}

\begin{figure}[htbp]
\includegraphics[width=0.55\columnwidth]{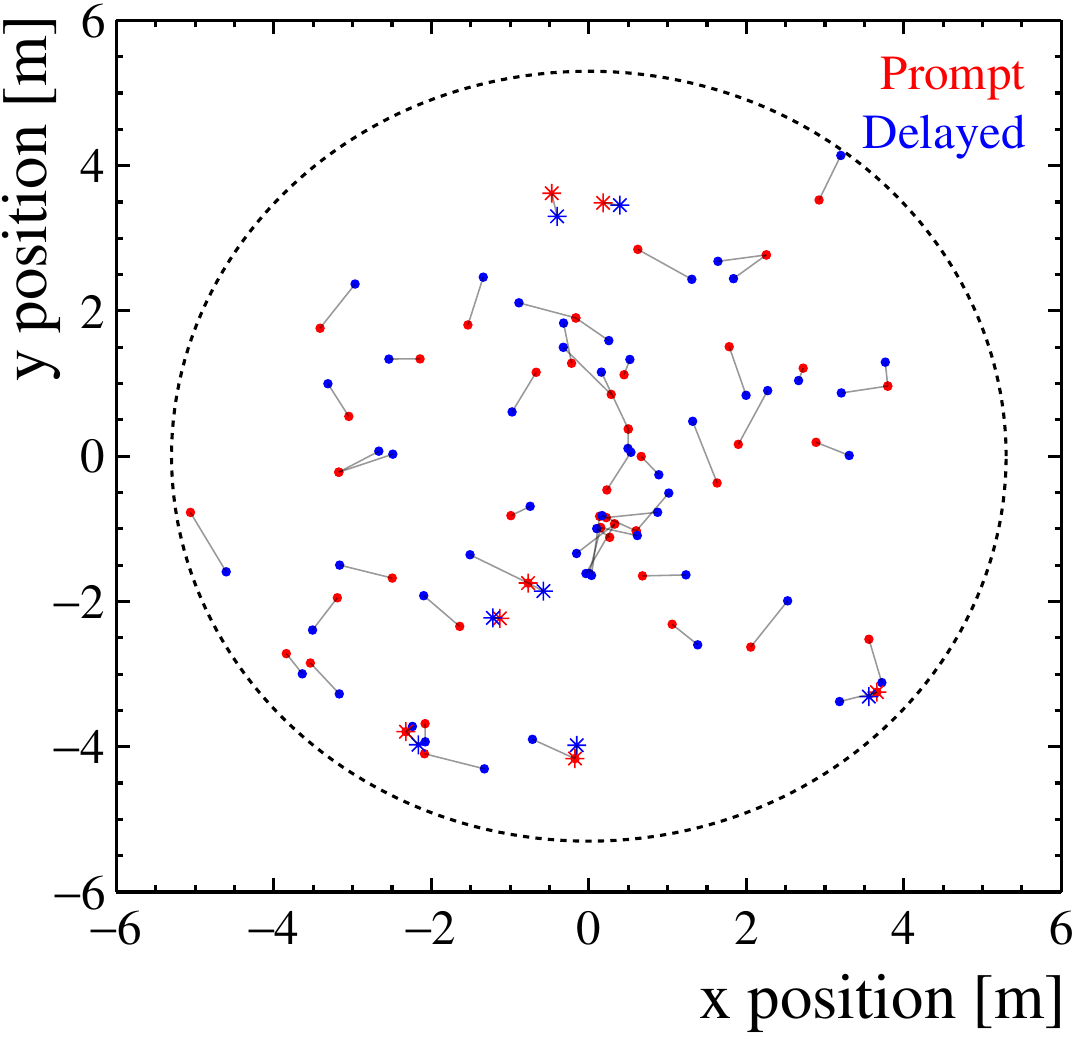}
\caption{\label{fig_x_vs_y}  x and y distribution of the 60 events. Red and blue points indicate prompt and decayed events respectively. The star points are ${}^{13}\text{C}$ signal-like events with LR $>$ 0. Solid circles have LR  $<$ 0 and are background-like events. The fiducial volume of R $<$ 5.3 m is shown as a black dashed line.} 
\end{figure}

\begin{figure}[htbp]
\centering
\includegraphics[width=0.55\columnwidth]{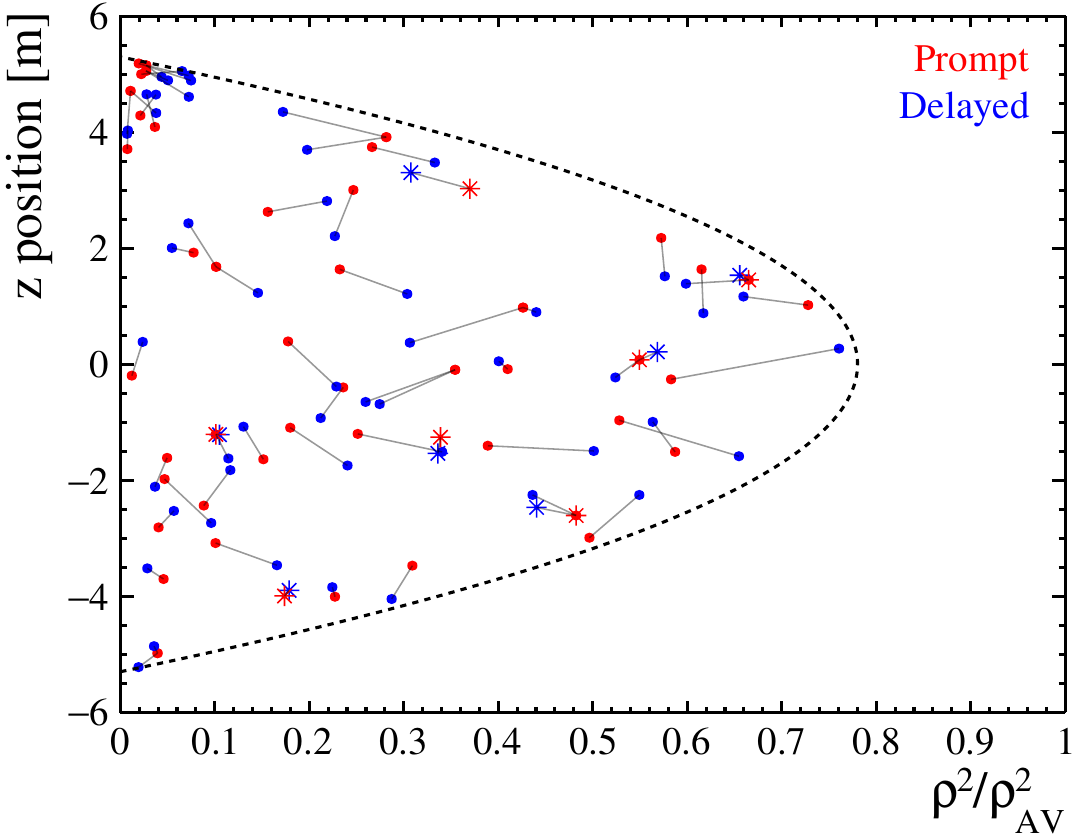}
\caption{\label{fig_rho_vs_z}z vs area-weighted $\rho^{2}$ distribution of the 60 events where $\rho =\sqrt{x^{2}+y^{2}}$ and $\rho_{av}$ = 6.0 m. Red and blue points indicate prompt and decayed events respectively. The star points are ${}^{13}\text{C}$ signal-like events with LR $>$ 0. Solid circles have LR  $<$ 0 and are background-like events. The fiducial volume of R $<$ 5.3 m is shown as a black dashed line.}
\end{figure}

\begin{figure}[htbp]
\includegraphics[width=0.7\columnwidth]{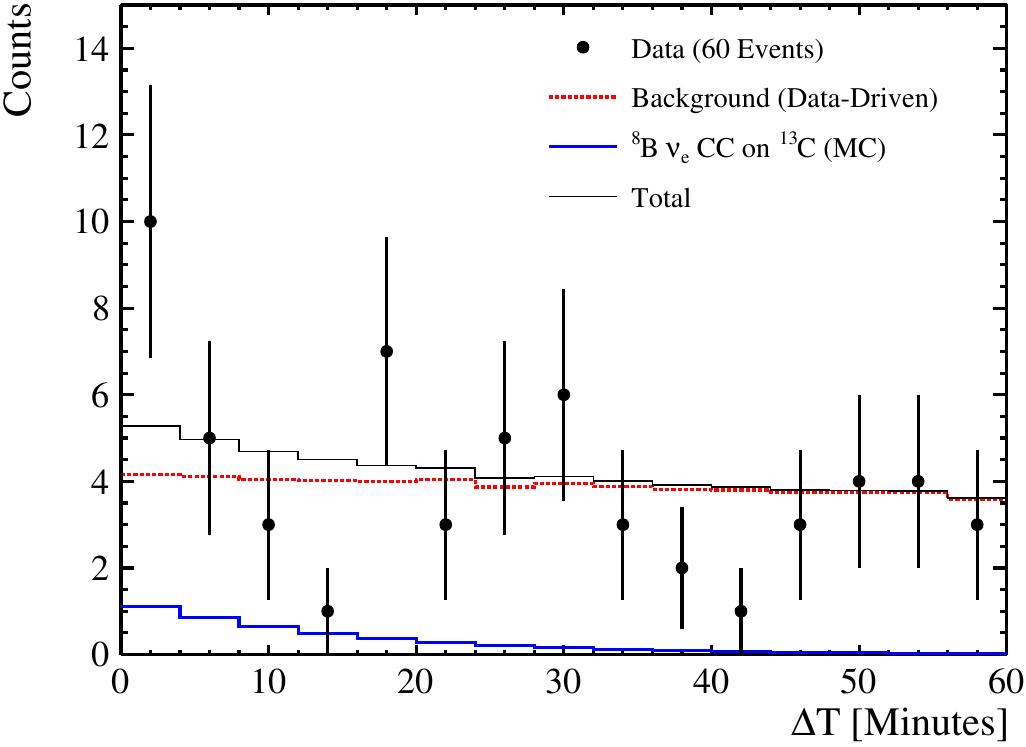}
\caption{\label{fig:delayed_pdfs} Time ($\Delta T$) between prompt and delayed events. The $^{13}$C signal is shown as a blue line from Monte Carlo simulations. The background distribution, derived using a data-driven method, is shown as a red dashed line. The normalisation has been scaled to the expected rate. The sum of the signal and background is shown as a thinner black line. The 60 observed data events are displayed as black points. }
\end{figure}

\begin{figure}[htbp]
\includegraphics[width=0.7\columnwidth]{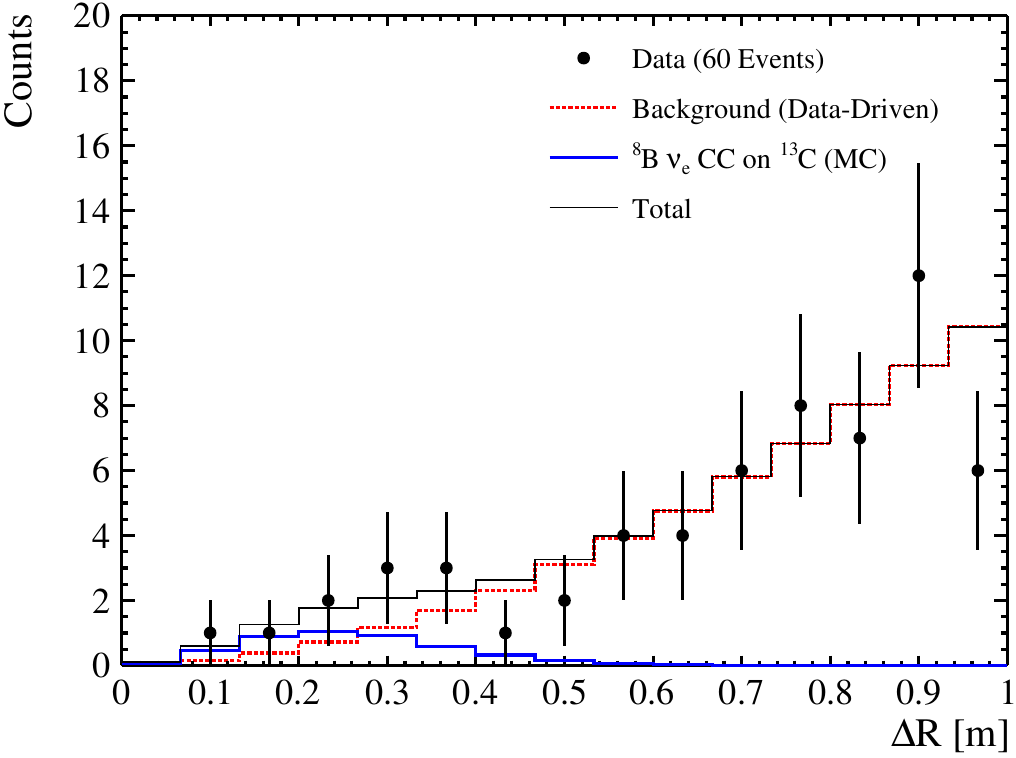}
\caption{\label{fig:delayed_pdfs} Distance ($\Delta R$) between the 3D reconstructed position of the prompt and delayed events. The $^{13}$C signal is shown as a blue line from Monte Carlo simulations. The background distribution, derived using a data-driven method, is shown as a red dashed line. The normalisation has been scaled to the expected rate. The sum of the signal and background is shown as a thinner black line. The 60 observed data events are displayed as black points. }
\end{figure}

\begin{figure}[htbp]
\includegraphics[width=0.7\columnwidth]{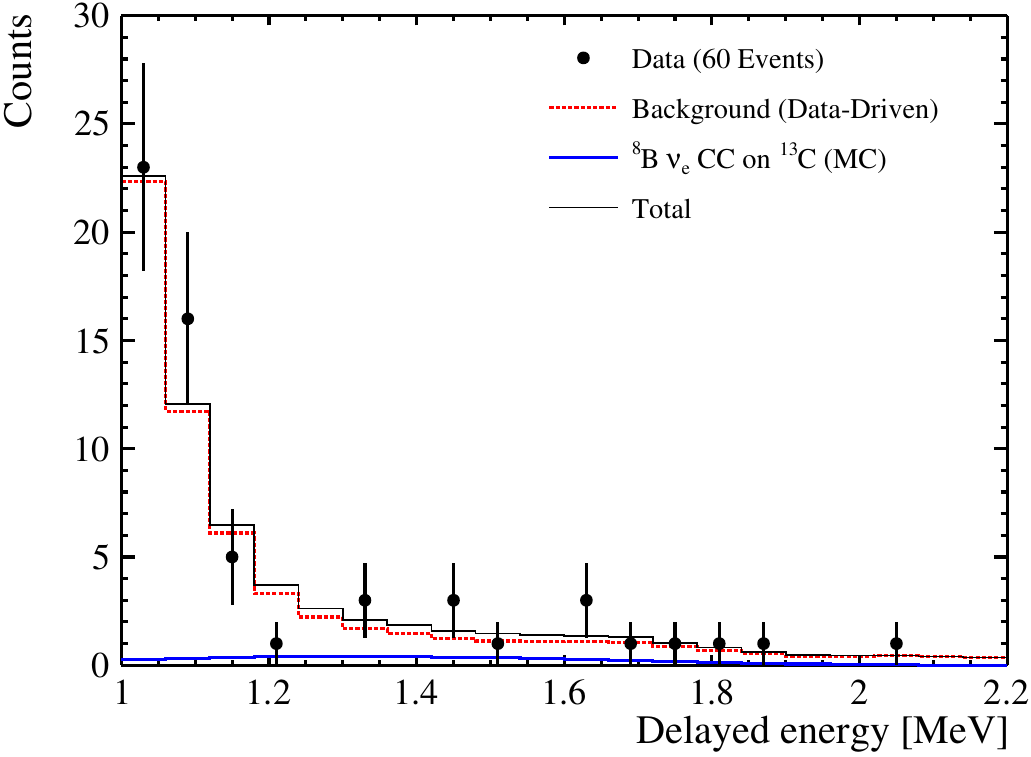}
\caption{\label{fig:delayed_pdfs} The energy of all the delayed candidate events. The $^{13}$C signal is shown as a blue line from Monte Carlo simulations. The background distribution, derived using a data-driven method, is shown as a red dashed line. The normalisation has been scaled to the expected rate. The sum of the signal and background is shown as a thinner black line. The 60 observed data events are displayed as black points. }
\end{figure}